\documentclass{aa}

\usepackage{color}
\usepackage{ctable}
\usepackage{twoopt}
\usepackage{multirow}
\usepackage{amsmath,amssymb}
\usepackage{threeparttable}
\usepackage[breaklinks=true]{hyperref} 
\bibpunct{(}{)}{;}{a}{}{,} 
\makeatletter
\newcommandtwoopt{\citeads}[3][][]{\href{http://adsabs.harvard.edu/abs/#3}%
{\def\hyper@linkstart##1##2{}%
\let\hyper@linkend\@empty\citealp[#1][#2]{#3}}}
\newcommandtwoopt{\citepads}[3][][]{\href{http://adsabs.harvard.edu/abs/#3}%
{\def\hyper@linkstart##1##2{}%
\let\hyper@linkend\@empty\citep[#1][#2]{#3}}}
\newcommandtwoopt{\citetads}[3][][]{\href{http://adsabs.harvard.edu/abs/#3}%
{\def\hyper@linkstart##1##2{}%
\let\hyper@linkend\@empty\citet[#1][#2]{#3}}}
\newcommandtwoopt{\citeyearads}[3][][]%
{\href{http://adsabs.harvard.edu/abs/#3}
{\def\hyper@linkstart##1##2{}%
\let\hyper@linkend\@empty\citeyear[#1][#2]{#3}}}
\makeatother
\usepackage{hyperref}

\usepackage{txfonts}
\usepackage{longtable}
\usepackage{rotating}
\usepackage{graphicx}
\usepackage{graphics}
\usepackage{psfrag}
\usepackage{amssymb}
\bibpunct{(}{)}{;}{a}{}{,}
\usepackage{color}
\usepackage{ctable}
\usepackage{threeparttable}
\usepackage{ mathrsfs }

\def\Teff{$T_{\mathrm{eff}}$}
\def\logg{\ensuremath{\log g}}

\def\vmac{$\upsilon_{\mathrm{macro}}$}
\def\vsini{\ensuremath{{\upsilon}\sin i}}

\def\Teff{\ensuremath{T_{\mathrm{eff}}}}
\def\logg{\ensuremath{\log g}}

\def\vmac{$\upsilon_{\mathrm{macro}}$}
\def\vsini{\ensuremath{{\upsilon}\sin i}}

\def\Pt{\ensuremath{\rm P_{\mathrm{turb}}\,}}
\def\Ptmax{\ensuremath{\rm P_{\mathrm{turb}}^{\mathrm{max}}\,}}

\def\Msun{\ensuremath{\rm \,M_\odot\,}}

\def\logLl{\rm log(L/L_{\odot})}
\def\vmac{$\upsilon_{\mathrm{macro}}$}
\def\vmacth{$\overline{\upsilon}_{\mathrm{macro}}$}

\begin{document}

\title{Metallicity dependence of turbulent pressure and macroturbulence in stellar envelopes}

\author{L. Grassitelli$^1$, L. Fossati$^{2,1}$, N. Langer$^1$, S. Sim\'on-D\'iaz$^{3,4}$, N. Castro$^{5,1}$, D. Sanyal$^1$}
\institute{Argelander-Institut f\"ur Astronomie der Universit\"at Bonn, Auf dem H\"ugel 71, 53121, Bonn, Germany \and
Space Research Institute, Austrian Academy of Sciences, Schmiedlstrasse                 6, A-8042 Graz, Austria \and
Instituto de Astrof\'{i}sica de Canarias, 38200 La Laguna, Tenerife, Spain \and 
Departamento de Astrof\'isica, Universidad de La Laguna, Avda. Astrof\'isico Francisco S\'anchez, s/n, 38200 La Laguna, Tenerife, Spain \and Department of Astronomy, University of Michigan, 1085 S. University Avenue, Ann Arbor, MI 48109-1107, USA
}

\abstract{
Macroturbulence, introduced as a fudge to reproduce the width and shape of stellar absorption lines, reflects gas motions in stellar atmospheres. While in cool stars, it is thought to be caused by convection zones immediately beneath the stellar surface, the origin of macroturbulence in hot stars is still under discussion. Recent works established a correlation between the turbulent-to-total pressure ratio inside the envelope of stellar models and the macroturbulent velocities observed in corresponding Galactic stars. To probe this connection further, we evaluated the turbulent pressure that arises in the envelope convective zones of stellar models in the mass range 1--125 \Msun\ based on the mixing-length theory and computed for metallicities of the Large and Small Magellanic Cloud. We find that the turbulent pressure contributions in models with these metallicities located in the hot high-luminosity part of the Hertzsprung-Russel (HR) diagram is lower than in similar models with solar metallicity, whereas the turbulent pressure in low-metallicity models populating the cool part of the HR-diagram is not reduced. Based on our models, we find that the currently available observations of hot massive stars in the Magellanic Clouds appear to support a connection between macroturbulence and the turbulent pressure in stellar envelopes. Multidimensional simulations of sub-surface convection zones and a larger number of high-quality observations are necessary to test this idea more rigorously.}

\keywords{star: massive mass - convection zone - turbulence - pulsations - Magellanic Clouds}
\titlerunning{Turbulent pressure and macroturbulence broadening}
\authorrunning{L. Grassitelli et al.}
\maketitle

\section{Introduction}

The Large and Small Magellanic Clouds (LMC and SMC) are two dwarf satellite galaxies of the Milky Way (MW). The close distance of these two galaxies 
 makes it possible to obtain high-quality spectra from individual stars \citep[see, e.g.,][]{2011Evans,2014Bestenlehner}. This makes them great laboratories in which we can investigate several open questions in stellar physics, such as stellar evolution in different environments \citep{2013Yusof,2014Schneider,2015Kohler}, the effects of different opacities on the structure of stars and its implications \citep{1997Sasselov,2006Keller,2009Cantiello}, and the relation between the ambient metallicity and powerful events such as long-duration gamma-ray bursts and superluminous supernovae \citep{2013Graham,2014Lunnan,2014Kozyreva}. From this point of view, their main difference compared to our Galactic neighborhood is their different metallicity (about half and a fifth of the solar metallicity for the LMC and SMC, respectively). The different chemical composition affects the opacity profile within the stars, of importance mostly in the sub-surface envelopes where temperatures are relatively low, such that full ionization cannot be sustained. Especially the recombination of the iron-group elements at about 200\,000\,K plays an important role in defining the structure of the outer layers of stars close to the Eddington limit, which is strongly influenced by the metallicity of the star \citep{2012Langer,2012Goetze,2015Jiang,2015Sanyal}. The corresponding opacity bump, together with the opacity bumps due to hydrogen and helium recombination, often induces large-scale turbulent motions of material, or in other words, convection, in the sub-surface layers \citep{1993Stothers,2009Cantiello,2015GrassitelliA}. 

 In the context of Galactic massive OB stars, \citet{2015GrassitelliA} recently found a strong correlation between the strength of the pressure arising from turbulent motion in the iron convective zone (FeCZ) of their stellar models and the observational appearance of an additional broadening of the absorption lines in the spectra of corresponding stars. This additional broadening is commonly called macroturbulence  \citep[][and reference therein]{2011SimonDiaz,2015SimonDiaz,2012SimonDiaz}, and it is thought to be caused by large-scale surface motion. The introduction of this additional broadening of the spectral lines has been motivated and supported by the lack of apparent slow rotators and sharp-lined stars in the Galactic O-B stars \citep{2004Howarth,2014SimonDiaz}. Although the association between this additional broadening and motion in the surface layers appears straightforward \citep{1956Slettebak,1977Conti}, the origin of this velocity field is not yet fully understood.  

However, macroturbulence is not only present in the spectra of massive hot main-sequence stars and supergiants. The same type of broadening is also present in other regions of the Hertzprung-Russel diagram \citep[HR diagram, ][]{fossati2009,2014Doyle}. \citet{2015GrassitelliB} found a similar correlation between turbulent pressure (\Pt) and macroturbulent velocities (\vmac) in the context of late-type low- and intermediate-mass stars. In this case they considered the effects of turbulence in the hydrogen convective zone (HCZ), suggesting that a common mechanism across the HR-diagram might induce the surface velocity field observed as macroturbulent broadening. This mechanism was suggested to rely on the excitation of oscillations, as \citet{2009Aerts} showed that a high number of high-order nonradial g-mode pulsations can collectively reproduce the effects on the spectral lines that are typical for the macroturbulent broadening. 

Based on this, the relation between turbulent pressure and macroturbulent velocities has been interpreted as due to the broad spectrum of pulsations excited by the turbulent pressure fluctuations at the percent level \citep{2015GrassitelliA}. This dynamical mechanism to excite pulsations is expected to act effectively in the sub-surface convective zones in which convection is not an efficient mechanism to transport energy. This leads to a turbulent flow with high convective velocities and consequently high Reynolds numbers \citep{1991Canuto,2015Arnett}.    
  
We investigate here the effects of the metallicity on the strength of the turbulent pressure in the sub-surface convective zones across the HR-diagram.

\section{Method}

We computed stellar models using BEC, a dedicated implicit Lagrangian one-dimensional hydrodynamic stellar evolution code \citep{2000Heger,2005Petrovic,2006Yoon,2011Brott,2015Kohler}. It includes current physics and treats convection following the nonadiabatic standard mixing-length theory \citep[MLT,][]{1958Vitense}, with a mixing-length parameter $\rm \alpha=1.5$ \citep{2011Brott}. The Rosseland mean opacity of stellar matter is computed with data from the OPAL opacity tables from \citet{1996Iglesias} in the high-temperature regime and from \citet{1994Alexander} in the low-temperature regime (i.e., below log$(T/K)=3.75$).
Our calculations include mass-loss by stellar wind according to the prescription by \citet{2001Vink}, while other physical parameters have been chosen as in \citet{2011Brott}.   

We estimated the strength of turbulent motion in the partial ionization zones by considering the turbulent pressure \Pt as in Grassitelli et al.\citep[2015a, see also ][]{1991Canuto,1997Jiang,2003Stothers,2009Maeder}:
\begin{equation}\label{Eq.pturb}
P_{turb}= \zeta \rho \upsilon_c^2 \quad ,
\end{equation}
where $\rho$ is the local density, $\zeta$ is a parameter chosen to be $\zeta=1/3$ for isotropic turbulence \citep{2003Stothers,2009Maeder}, and $\upsilon_c$ is the local convective velocity. 

The validity regime of the MLT is the subsonic regime, which
is when the convective eddies move at a speed significantly lower than the local sound speed. The MLT does not take dissipation in case of transonic convective eddies into account and assumes pressure equilibrium with the medium \citep{1958Vitense}. Therefore, given that supersonic convection is unphysical, we limited $\upsilon_c$ to the local isothermal sound speed ($c_s$) with the criteria \citep{2015GrassitelliA,2015GrassitelliB}
\begin{equation}\label{soundcriteria}
\upsilon_c\leq c_s \quad ,\quad c_s^2= \frac{k_B T}{\mu m_H}= \frac{P_{gas}}{\rho} \quad ,
\end{equation}
where $T$ is the local temperature, $k_b$ is the Boltzmann constant, $\mu$ is the mean molecular weight, and $m_H$ is the hydrogen mass. 
Consequently, the turbulent pressure can at most be equal to a fraction $\zeta$ of the gas pressure.

We did not directly include the effects of turbulent pressure and turbulent energy density in our calculations, estimating the turbulent pressure \textup{{\it a posteriori} }through Eq.\ref{Eq.pturb}, since \citet{2015GrassitelliA,2015GrassitelliB} showed that the structural differences as a consequence of including turbulent pressure and energy density are small. We expect these effects to be even smaller in low-metallicity environments.

\section{Results}

\begin{figure}
\resizebox{\hsize}{!}{\includegraphics{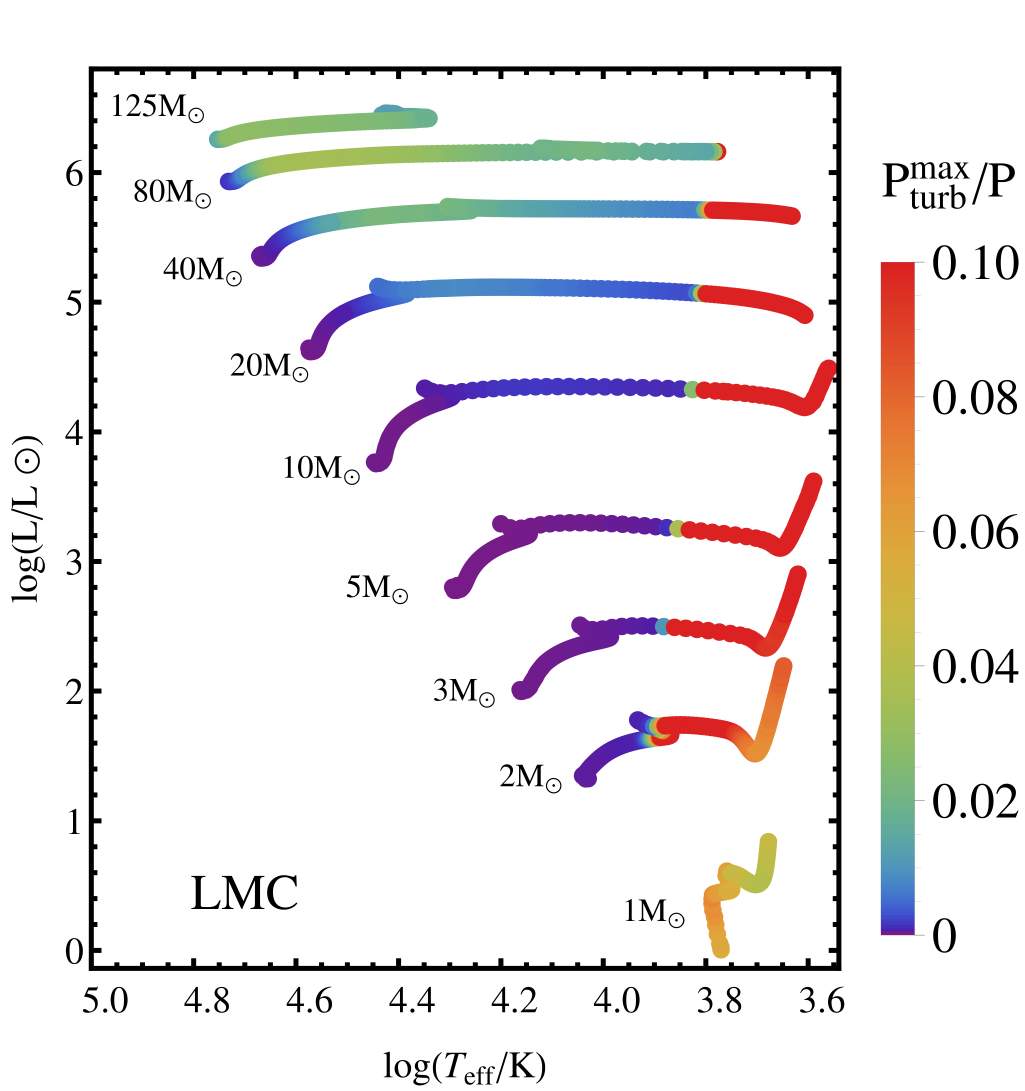}}\caption{HR-diagram showing our stellar evolutionary tracks with LMC metallicity. Each dot indicates a stellar model color-coded according to the highest ratio of turbulent-to-total pressure encountered within the stellar model (see color bar on the right). The numbers close to the tracks indicate the zero-age
main-sequence mass of the models. Models with \Ptmax/P between 10\% and 33\% are shown in red. The absolute maximum is found for the massive red supergiants.}\label{LMCtr}
\end{figure}

\begin{figure}
\resizebox{\hsize}{!}{\includegraphics{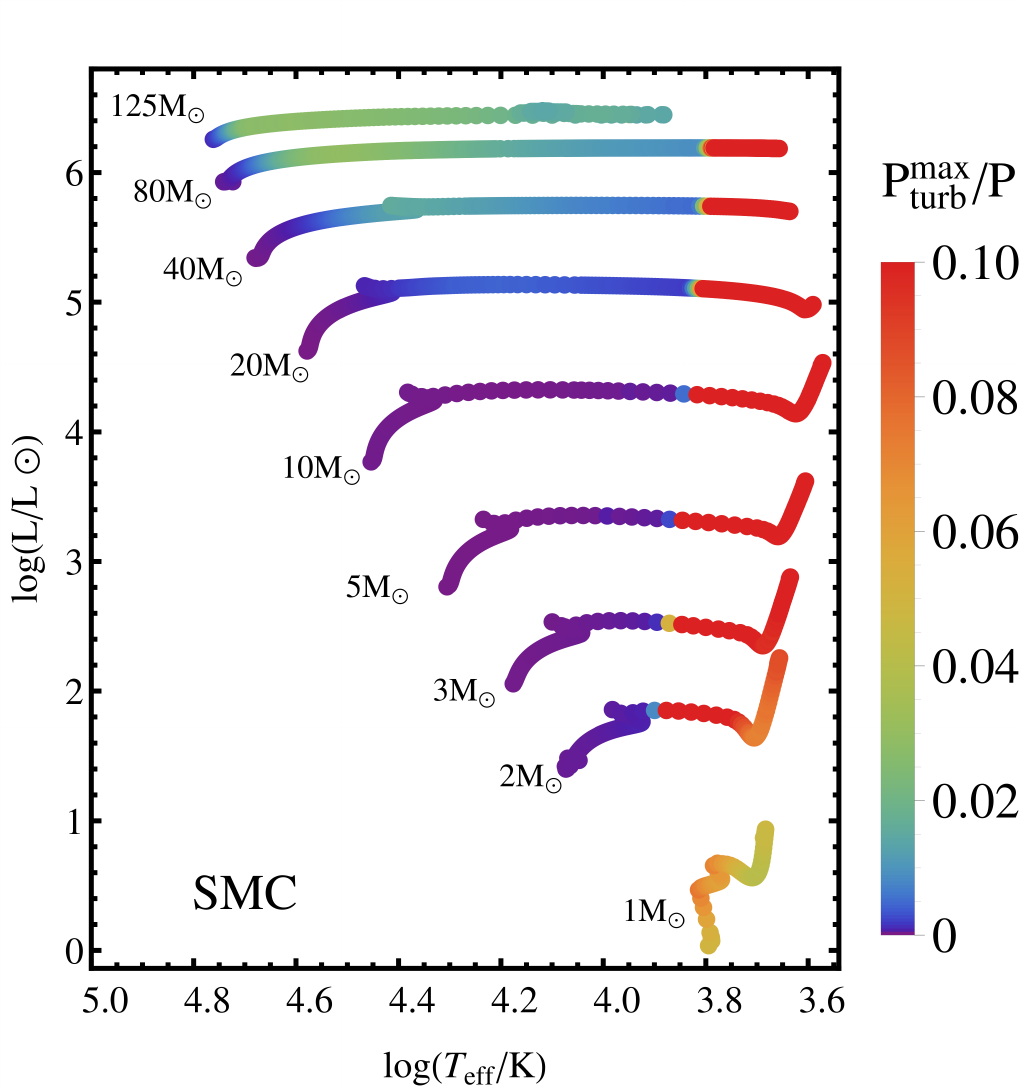}}\caption{Same as Fig.\ref{LMCtr}, but for SMC metallicity.}\label{SMCtr}
\end{figure}

We computed a set of stellar models and stellar evolutionary tracks for nonrotating main-sequence (MS) and post-MS stars with LMC and SMC metallicity (as defined in \citealt{2011Brott}) in the stellar mass range $\rm 1-125\,M_\odot$. The stellar evolutionary tracks in the HR-diagram are shown in Figs.\ref{LMCtr} and \ref{SMCtr}, with each stellar model color-coded according to the highest fraction of turbulent pressure within the stellar envelopes, \Ptmax/P.

\subsection{OB stellar models}

Considering at first the hotter stellar models (i.e., log(\Teff/K) $\gtrsim$ 4), the computed evolutionary tracks show that at low luminosities ($\logLl<4$) \Pt is small ($\Ptmax/P\ll1\%$). It tends to increase for higher luminosity, that is, in the most massive models, for both LMC and SMC metallicity, reaching values of \Ptmax/P  of about 3--4\% at $\logLl\approx6$. 

\begin{figure*}
\resizebox{\hsize}{!}{\includegraphics{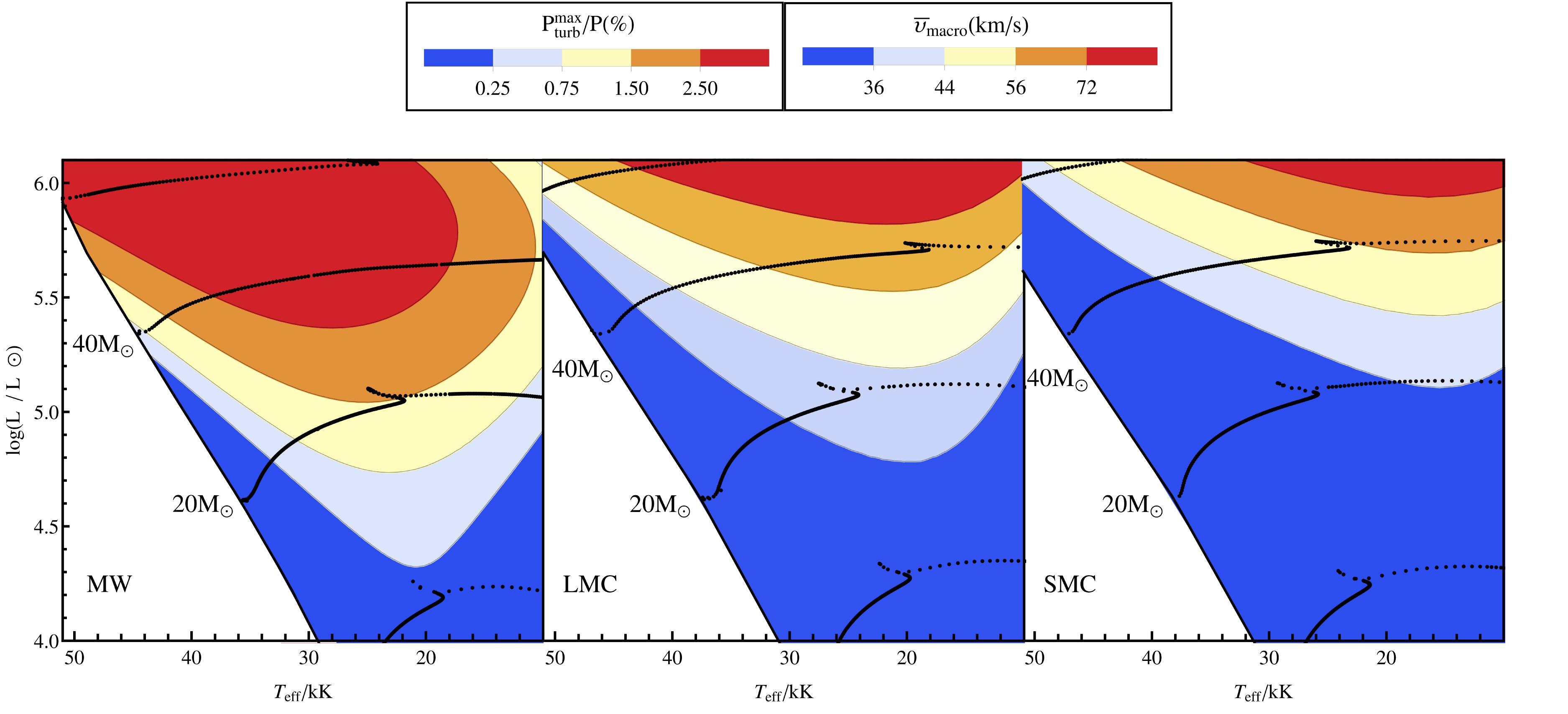}}\caption{HR diagrams color-coded as a function of the highest fraction of turbulent pressure as derived from a best
fit of the tracks in Fig. \ref{LMCtr} (central pannel, LMC metallicity), Fig. \ref{SMCtr} (right pannel, SMC metallicity), and from Fig. 1 of \citet{2015GrassitelliA} (left pannel, MW metallicity). The values of the expected macroturbulent velocities (\vmacth) given in the top bar were obtained by adopting the linear relation between the observationally derived macroturbulent velocities of a sample of Galactic OB stars and the corresponding highest fraction of turbulent pressure from Fig. 5 of \citet{2015GrassitelliA}.}\label{triplo}
\end{figure*}

When comparing the stellar tracks in Figs. \ref{LMCtr} and \ref{SMCtr} with the Galactic ones in Fig.1 of \citet{2015GrassitelliA}, it appears that the turbulent pressure for hot, massive stars is metallicity dependent. This can be seen more clearly in Fig.\ref{triplo}, where the top part of the HR-diagram is shown for the three different metallicities. The differently colored regions in the three panels (result of a best fit of the tracks) show that the same highest contribution of turbulent pressure is located at higher luminosities as the metallicity decreases. 
Thus there is a trend of a larger turbulent pressure fraction at higher metallicity, at either fixed luminosity or fixed initial stellar mass. This can be explained when considering that the iron opacity bump is less pronounced at metallicities lower than the Galactic one. Consequently, the envelope of the stars inflates less strongly for an LMC metallicity, and even weaker for an SMC metallicity, compared to stars of MW metallicity (Sanyal et al. in prep.). This in turn leads to higher densities in the FeCZ, and thus, together with the lower opacities, to more efficient convection. 
Efficient convection implies low degrees of super-adiabaticity in the convective layers (and vice versa), which in turn implies  low convective velocities and thus low pressure arising from the turbulent motion computed through Eq.\ref{Eq.pturb}. 

The ratio \Pt/P can be rewritten as

\begin{equation}
\frac{P_{turb}}{P} = \frac{\zeta \rho \upsilon_c^2}{P} = \frac{\zeta P_{gas} (\upsilon_c/c_s)^2}{P}  \quad ,
\end{equation}    
and therefore
\begin{equation}\label{machbeta}
\frac{P_{turb}}{P} = \zeta \beta M_c^2  \quad ,
\end{equation}    
where $M_c$ is the Mach number defined as the ratio $v_c/c_s$, and $\beta$ is the local ratio of the gas-to-total pressure. In the hot, luminous part of the HR diagram, Eq.\ref{machbeta} leads to a local maximum of \Ptmax/P at $\logLl \approx 6$ in both Figs.\ref{LMCtr} and \ref{SMCtr}. The reason is that  at first the Mach numbers increase moving from $\logLl=0$ to $\logLl \approx 6$, until the convective velocities derived from the MLT reach the sound speed. As the convective velocities become transonic, Eq.\ref{soundcriteria} limits the Mach number to 1, and therefore the only variable left in Eq.\ref{machbeta} is $\beta,$ which, as already mentioned, decreases toward higher luminosities, consequently leading to a local maximum for \Ptmax/P (see Fig.\ref{ptL} in Sect. 4). Equation \ref{machbeta} also implies that for transonic convective velocities and a gas-pressure-dominated convective zone, the highest value for the ratio \Ptmax/P is one third.

\subsection{Late-type main-sequence stars and red giant models}

\begin{figure}\vspace{0.5cm}
\resizebox{\hsize}{!}{\includegraphics{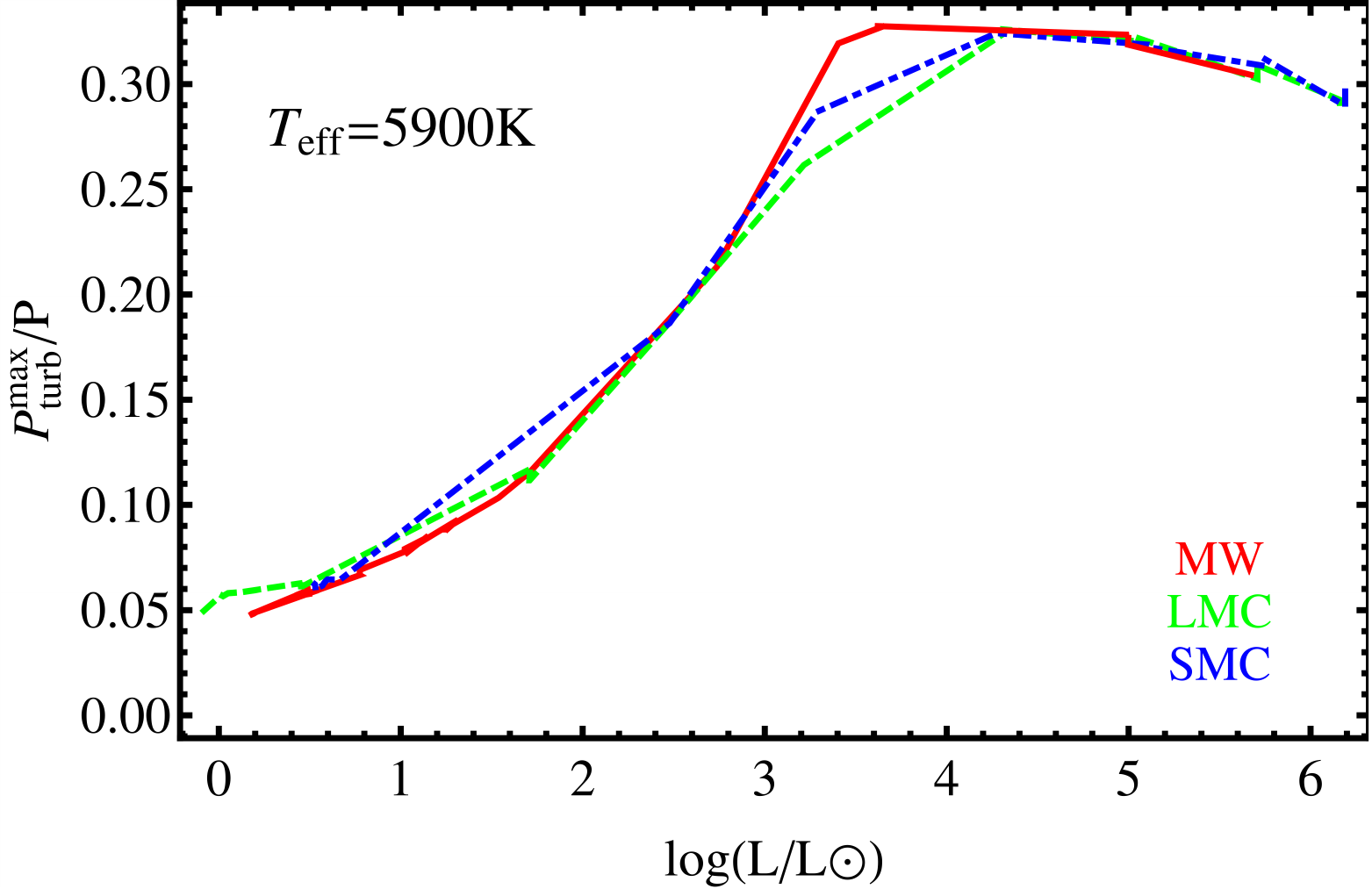}}\caption{Highest ratio of turbulent-to-total pressure in stellar models at the effective temperature of \Teff=5900K, with MW (red, continuous line), LMC (green, dashed line), and SMC (blue, dot-dashed line) metallicity as a function of their luminosity. The MW models are taken from \citet{2015GrassitelliA}. }\label{redg}
\end{figure}

Given that the metallicity is the fraction of mass of a star that is not in hydrogen or helium, we do not expect significant differences in comparison to the Galactic case for the convective zones arising at the temperatures where H and He recombine \citep[where the helium convective zone is generally not expected to initiate transonic convective motion; ][]{2009Cantiello,2015GrassitelliA,2015GrassitelliB}. This is the case for the cool stars (i.e., log(\Teff/K) $\lesssim$ 4) in Figs.\ref{LMCtr} and \ref{SMCtr}. Hydrogen recombines at a temperature of $\approx 8\,000\,$K, leading to a convective zone very close to the surface. 
As Figs.\ref{LMCtr} and \ref{SMCtr} show, the HCZ leads to an almost vertical band of high contribution from the turbulent pressure in the HR diagram when the temperature for the recombination of hydrogen is met within the star, very similar to the one found by \citet{2015GrassitelliB}. This band stretches from the zero-age
main-sequence of the models with about $\rm \approx 1\Msun$
up to the most luminous models.
While moving toward higher luminosities, the band widens and its sharp hot edge moves toward slightly cooler temperatures (see Figs.\ref{LMCtr} and \ref{SMCtr}). 

 The fraction of turbulent pressure arising in this convective zone is very high, especially in the more massive models. This is shown in Fig.\ref{redg}, where the fraction of turbulent pressure is shown to increase as a function of luminosity in the cool part of the HR diagram (at a fixed temperature of log(\Teff/K)$\approx3.77$). The fraction of turbulent pressure starts from $\approx 5\%$ at solar luminosity, increasing as the luminosity (and thus the mass) of our models increases, independent of the metallicity. The increase continues until $\logLl \approx 3.5$, where $\Pt/P$ reaches $\approx 33\%$ in  all the sets of models.  Based on the discussion in Sect.3.1, this value is the highest possible value because the sound speed is limited, as shown in Eq.\ref{soundcriteria}. At even higher luminosities, the fraction of turbulent pressure therefore continues at approximately one-third of the total pressure
and is only mildly affected by the decrease in gas pressure fraction. 

In this class of objects the high fraction of turbulent pressure it is not only due to the convective velocities approaching the sound speed (a characteristic shared with the hot luminous stellar models), but also to the outer layers of the cool stars, which
are gas-pressure dominated. The different $\beta$-values in the FeCZ of the massive stellar models (0.1 or lower) and in the HCZ (close to 1) explain the differences in the contribution from \Pt in the cool and in the hot luminous parts of the HR-diagrams (see Eq.\ref{machbeta}).

 Stellar models in the low-luminosity low-temperature corner of the HR-diagrams instead show a fractional contribution from \Ptmax of a few percent, which is lower than the value found in the slightly hotter models. After hydrogen has recombined within the hydrostatic structure of the models (shown by the hot edge of the bands), a maximum for the ratio \Ptmax/P is achieved throughout the evolution of a stellar model, followed by a gradual decrease \citep{2015GrassitelliB}. The reason is that as stellar models evolve, the HCZ moves to higher optical depths, leading to more efficient convection and therefore lower Mach numbers associated with the convective eddies.       

\subsection{Evolution at 40\Msun}

\begin{figure}
\resizebox{\hsize}{!}{\includegraphics{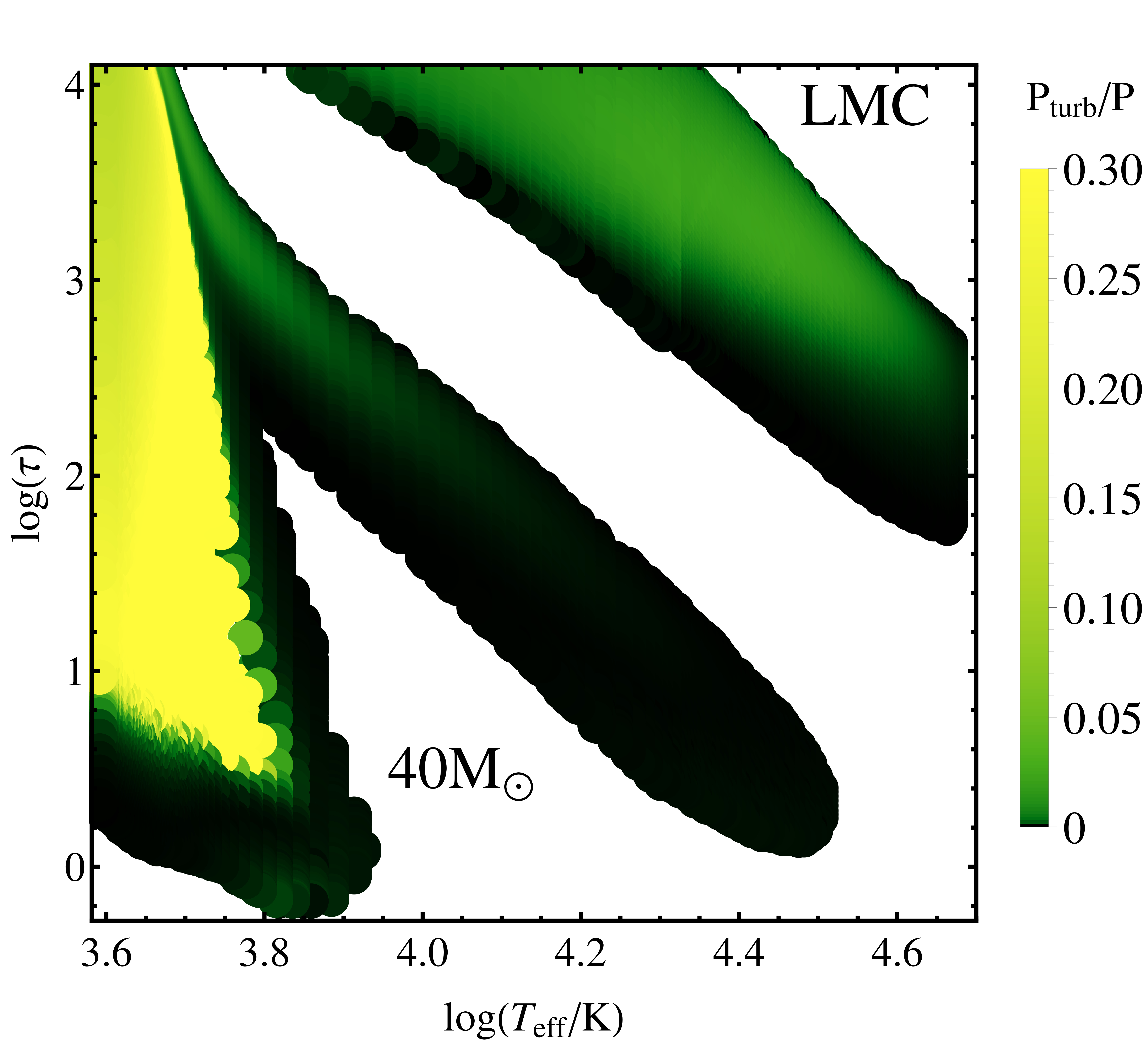}}\caption{Ratio of turbulent-to-total pressure (color-coded) as a function of effective temperature and optical depth throughout the evolution of our LMC 40\,\Msun model. The FeCZ is visible in the upper right corner, the HeCZ stretches across the diagram, and the HCZ appears for log(\Teff/K) below $\sim$3.9. The extended white regions are layers that are radiatively stable.}\label{LMCtau}
\end{figure}

\begin{figure}
\resizebox{\hsize}{!}{\includegraphics{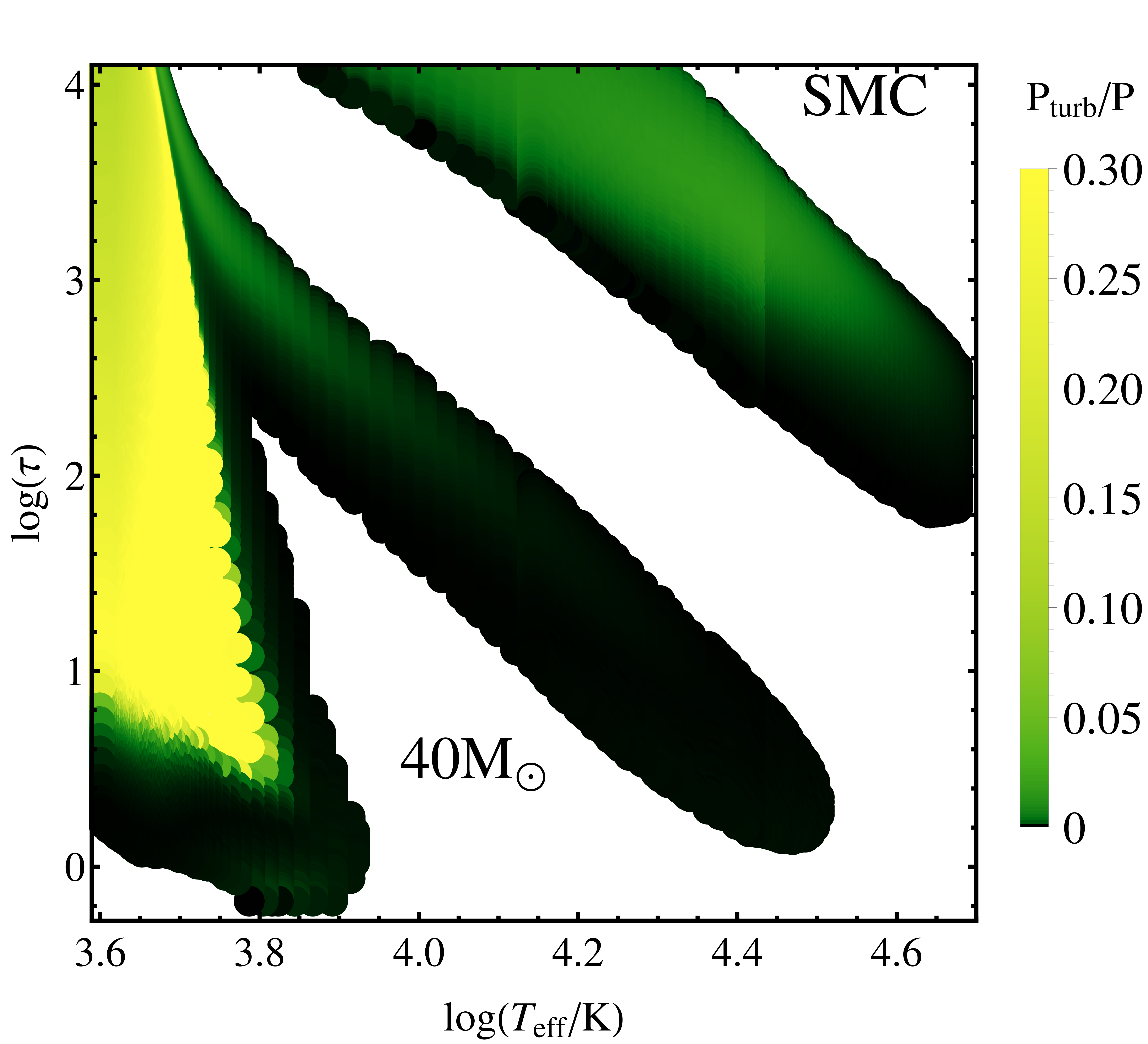}}\caption{Same as Fig.\ref{LMCtau}, but with SMC metallicity.}\label{SMCtau}
\end{figure}

In Figs.\ref{LMCtau} and \ref{SMCtau} the outer layers of our 40\,\Msun models for LMC and SMC metallicity are shown throughout their evolution as a function of log(\Teff/K) and logarithmic optical depth log($\tau$). When the two models are close to the zero-age main-sequence, only the FeCZ is present in an optical depth range log($\tau$)$\approx2-3$, showing in both cases a turbulent-to-total pressure contribution of a few percent. As the models evolve toward cooler effective temperatures, the FeCZ moves deeper inside the star up to an optical depth of log($\tau$)$\approx4$, and the temperatures become low enough for the HeCZ to arise. In this convective zone, however, the turbulent pressure constitutes, in general, only a small fraction of the local pressure ($\ll$ 1\,\%). In the temperature range log(\Teff/K)=3.8--3.9, the hydrogen opacity bump induces convection at the surface that increasingly extends as the stellar model enters the red supergiant phase. Figures \ref{LMCtau} and \ref{SMCtau} show how, below log(\Teff/K)=3.8, the HCZ extends from log($\tau)\approx0$ up to log($\tau)\gtrsim4$, with \Pt accounting for up to 33\% of the total pressure (limited by the criteria in Eq.\ref{soundcriteria}). 
 We also note that the turbulent pressure only depends on the physical conditions in the near-surface convective zones, but not on the evolutionary phase of the core of the stellar model \citep[see also ][]{2015GrassitelliB}.

\section{Observational diagnostics}

\begin{figure}
\resizebox{1.05\hsize}{!}{\includegraphics{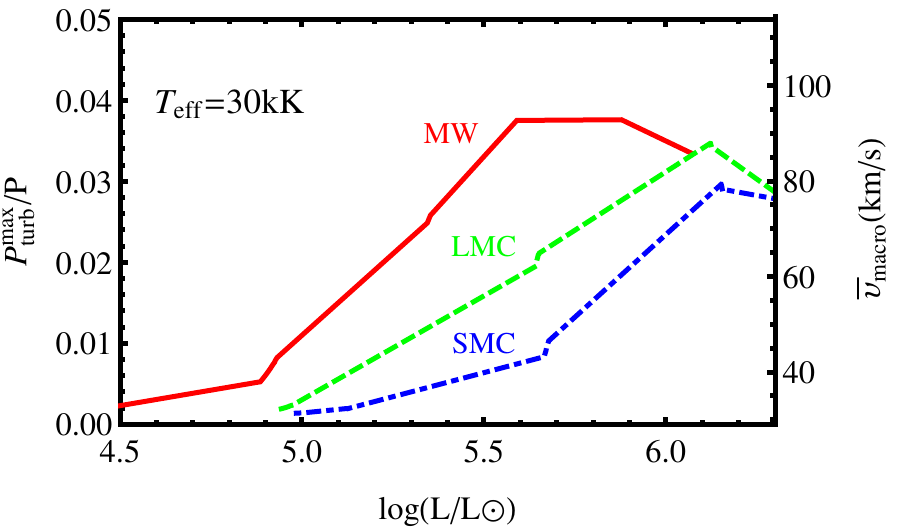}}\caption{Highest ratio of turbulent-to-total pressure in stellar models at the effective temperature of \Teff=30000K, with MW (red, continuous line), LMC (green, dashed line), and SMC (blue, dot-dashed line) metallicity as a function of their luminosity. The scale on the right indicates the expected macroturbulent velocities (\vmacth) derived from the linear fit in Fig. 5 of \citet{2015GrassitelliA}.}\label{ptL}
\end{figure}

\begin{figure}
\resizebox{1.05\hsize}{!}{\includegraphics{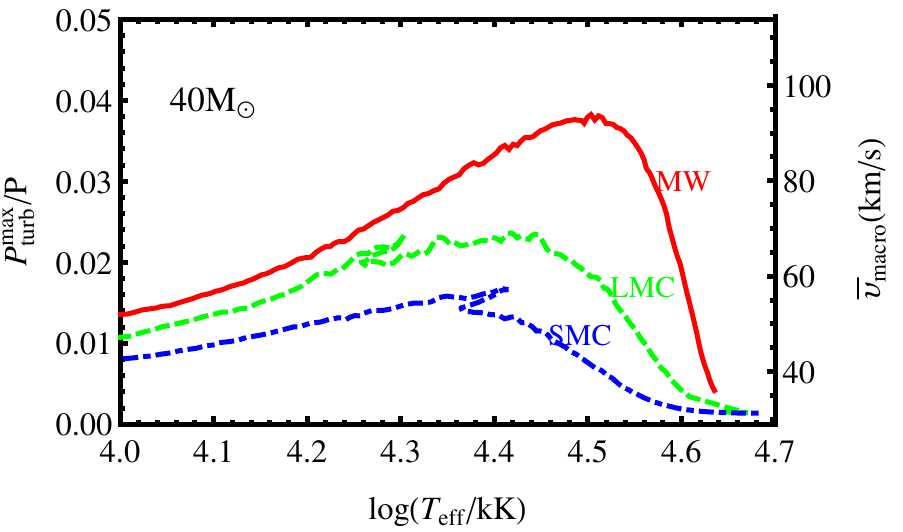}}\caption{Highest ratio of turbulent-to-total pressure in the envelopes of three 40\Msun stellar models with MW (red, continuous line), LMC (green, dashed line), and SMC (blue, dot-dashed line) metallicity as a function of effective temperature throughout part of their evolution. The scale on the right indicates the expected macroturbulent velocities (\vmacth) derived from the linear relation in Fig.
5 of \citet{2015GrassitelliA}. }\label{ptT}
\end{figure}

\citet{2015GrassitelliA,2015GrassitelliB} showed that the fractional contribution of the turbulent pressure across the HR-diagram can be related to macroturbulent broadening. This is the case for the Galactic OB stars \citep[cf. Fig. 5 by ][]{2015GrassitelliA}, for which a tight correlation  was found between the fraction of turbulent pressure and the observed macroturbulent velocities (\vmac). In these massive stars the high contribution from the turbulent pressure was encountered in the iron sub-surface convective zone at the base of the envelope. Similarly, a set of late-type low- and intermediate-mass stars showed a linear relation between the observed macroturbulent velocities and the estimated fraction of turbulent pressure \citep[cf. Fig. 3 by ][]{2015GrassitelliB}, arising in this case in the hydrogen convective zone located very close to the surface.    

Here, we investigate how macroturbulent broadening is expected to affect the atmosphere of stars with LMC and SMC metallicities. Unfortunately, unlike for the Galactic case \citep{2015SimonDiaz}, we lack a large sample of observationally derived macroturbulent velocity values for stars in the LMC and SMC, and therefore a direct statistical comparison is not possible at the moment. However, we compare our predictions to the macroturbulent velocities of a small sample of early B-type stars derived by \citet{2006Dufton} in Sect.\ref{obsmac}. To the best of our knowledge, this was the only sample of observationally derived macroturbulent velocities, together with \Teff\ and \logg, available in the literature.  
 
Assuming a similar correlation between \Ptmax and \vmac\ as found for the Galactic stars also for the Magellanic stars, we predict the macroturbulence to become stronger at higher luminosities compared to the Milky Way stars (see Fig.\ref{triplo}).
For example, in the LMC and SMC we expect to find macroturbulent velocities induced by turbulent pressure higher than 50 km/s only above luminosities of $\logLl\approx 5.3-5.5$ and $\logLl\approx 5.6-5.8$, respectively, while at solar metallicities we find this to occur starting from $\logLl\approx5$.

Figure \ref{ptL} shows the highest contribution from turbulent pressure as a function of luminosity and metallicity at a fixed effective temperature of \Teff=30\,kK, and macroturbulent velocities are associated adopting the linear fit\footnote{The linear fit relating \Ptmax and \vmac\ in fig.5 by \citet{2015GrassitelliA} writes \vmac$ \simeq 31+1664 \Ptmax$/P.} of the Galactic stars of \citet{2015GrassitelliA}. The fraction of turbulent pressure increases at higher luminosity, and consequently, the expected values for the macroturbulent velocities increase as a function of luminosity. Figure \ref{ptL} also shows the local maximum of the turbulent pressure discussed in Sect.3.1, which occurs due to the limitation of the convective velocities to the local sound speed (see Eqs.\ref{soundcriteria} and \ref{machbeta}).

We remark that it is unclear whether the linear correlation derived by \citet{2015GrassitelliA} for the Galactic OB stars applies to the stars of the Magellanic Clouds. This is mostly because the metallicity affects not only the derived values of the turbulent pressure, but also the whole structure of the envelope, potentially leading to a different correlation for the macroturbulent velocities. Moreover, the observationally derived macroturbulent velocities also include the spectral broadening due to microturbulence \citep{2014SimonDiaz}, and this is expected to be metallicity dependent \citep{2009Cantiello}.               
In any case, we expect stars in the Magellanic Clouds to show lower macroturbulent velocities in general.

 \citet{2009Penny} proposed, following indirect arguments, that macroturbulent broadening is metallicity dependent. These authors reached this conclusion by investigating the projected rotational velocity distributions of stars in the different environments, finding a cumulative distribution function that differed between the Galaxy and the Magellanic Clouds, as the LMC and SMC stars were showing lower \vsini\ values. This indirect argument provides empirical support for our scenario, as \citet{2009Penny} interpreted this difference as a consequence of lower macroturbulent velocities in LMC and SMC.
However, before reaching firm conclusions, our predictions need
to be compared with direct \vmac\ measurements of stars in the LMC and the SMC, ideally obtained following a similar approach as the one presented in the works by \citet{2011SimonDiaz,2015SimonDiaz} (i.e., considering a large sample of OB stars covering the whole region of the HRD above $\logLl \approx 3$).    

Figure \ref{ptT} shows the evolution of the highest fraction of turbulent pressure in the FeCZ as a function of effective temperature of three 40\Msun stellar models with different metallicities. This shows that in the early stages of the main-sequence evolution, the ratio \Ptmax/P tends to increase sharply for the MW metallicity (log($\Teff)\approx$ 4.6), slightly delayed and at lower effective temperature for the metallicities of the Magellanic Clouds (i.e., log($\Teff)\approx$ 4.5--4.6). A maximum is reached in the range log(\Teff/K)$\approx$4.4-4.5 (the exact value is a function of the metallicity and independent of the evolutionary phase), followed then by a decrease as the stellar models move toward low temperatures. Consequently, we expect the macroturbulent velocities to increase sharply at first in the O-type phase, followed by a decrease in the B-type phase until the red supergiant phase, where hydrogen recombines below the photosphere. This is in agreement with observations by \citet{2006Dufton}, \citet{2010Fraser}, and \citet{2010SimonDiaz}, who found a clear trend of decreasing macroturbulent velocities from \Teff=30\,kK to \Teff=10\,kK in a sample of B-type supergiants in the SMC and the MW. 

On the other side of the HR diagram, that is, in the low-temperature low-luminosity part in Figs.\ref{LMCtr} and \ref{SMCtr}, \citet{2015GrassitelliB} found a linear relation between \Ptmax/P and \vmac\, independent of the evolutionary stage of the models, but in this case with the highest turbulent pressure fraction found in the HCZ and not the FeCZ. In a recent work, \citet{2016Kitiashvili} found that the sub-surface convective motion indeed generates surface granulations and velocity fields on the order of tens of km/s, in line with the suggestion by \citet{2015GrassitelliB}. This was done by performing 3D radiative hydrodynamic simulations of the envelopes of F-type stars. The agreement between the induced velocity field and the characteristic values of macroturbulence in these objects is another indication of the probable connection between the two phenomena.  
 Given that the metallicity is not expected nor found to significantly influence the conditions in the HCZ, we predict a similar situation to also occur in the low-metallicity environments. If the connection between finite-amplitude turbulent pressure fluctuations and the $\gamma$-Doradus pulsating stars is confirmed, we expect to find this type of pulsators also at metallicities other than the Galactic one. 

\subsection{Comparison to macroturbulence measurements}\label{obsmac}

We collected the observationally derived values for the macroturbulent velocity from \citet{2006Dufton} for 13 B-type supergiants in the SMC and, by locating them in the HR-diagram, we related the observed macroturbulent velocities to the highest fraction of turbulent pressure in our models. 
To do so, we took into account that these macroturbulent velocities have been derived using a Gaussian line profile instead of the radial-tangential line profile \citep{2005Gray,2014SimonDiaz} adopted in most of the recent literature. Therefore, to consistently compare the SMC stars to the Galactic ones, we used the strong empirical relation in \citet{2014SimonDiaz} between the macroturbulent velocities derived through Gaussian and radial-tangential profiles, which means that all the macroturbulent velocities derived by \citet{2006Dufton} have been increased by a factor 1/0.65 \citep[cf. Fig. 3 of ][]{2014SimonDiaz}. 

\begin{figure}
\resizebox{1.0\hsize}{!}{\includegraphics{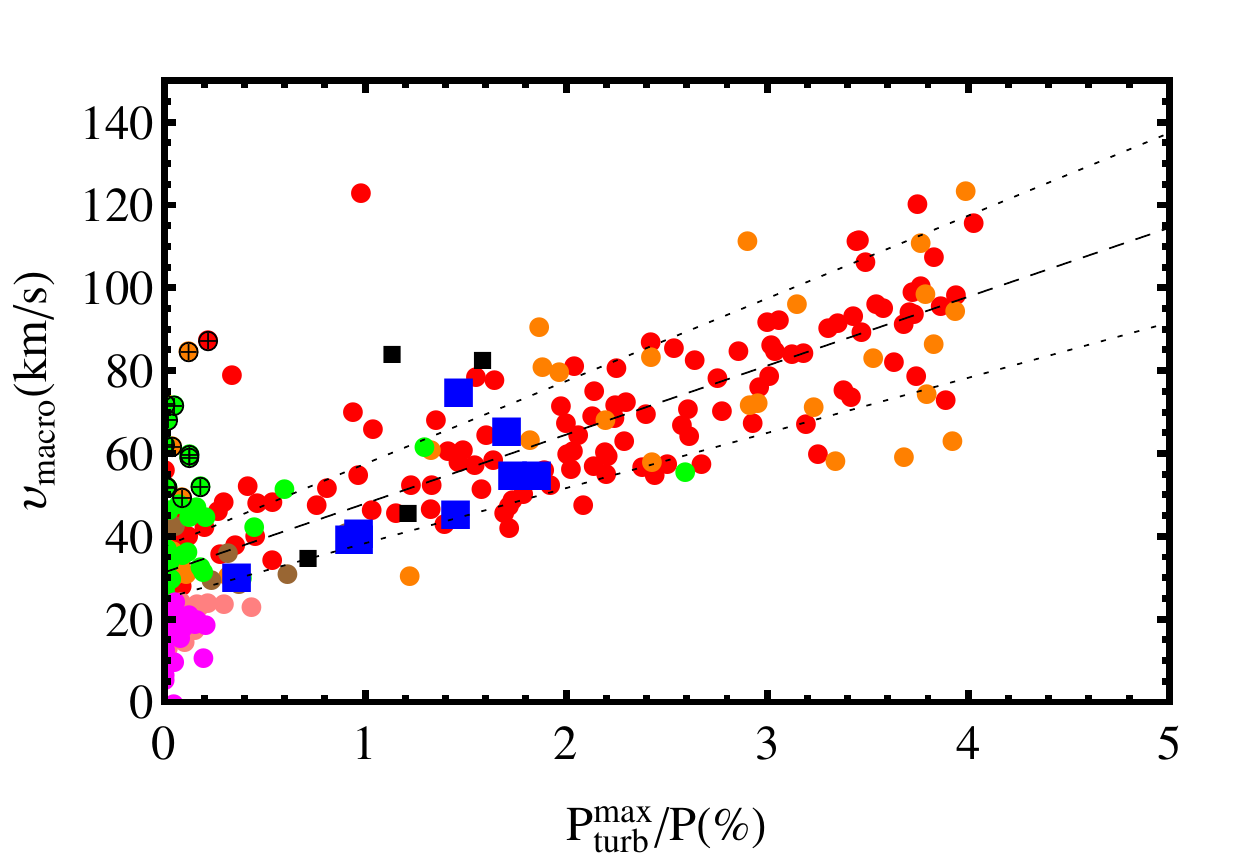}}
\caption{\vmac\ as a function of the highest ratio of turbulent-to-total pressure for the 13 B-type SMC stars (black and blue squares) in \citet{2006Dufton} compared to the set of Galactic OB stars (colored circles) in Fig.5 of Grassitelli et al.(2015a; for a description of the different colors, see \citealp{2015GrassitelliA} and \citealp{2015SimonDiaz}). Blue circles indicate the sub-sample of 9 stars used also in Fig.\ref{corL}. The dashed line indicates the linear best fit adopted to predict the theoretical macroturbulent velocities from Fig.5 by \citet{2015GrassitelliA}, while the dotted lines indicate standard error bars of $\approx 20 \%$ of \vmac.
}\label{correl}
\end{figure}

 We then directly compared the macroturbulent velocities as a function of \Ptmax/P for the stars in the SMC and the Galaxy. This is done in Fig.\ref{correl}, where the 13 B SMC stars are plotted superposed on the Galactic stars and their linear fit as shown in Fig. 5 of \citet{2015GrassitelliA}. As in \citet{2015GrassitelliA,2015GrassitelliB}, the two quantities appear linearly correlated, with a Spearman rank correlation coefficient of 0.74. Moreover, in Fig.\ref{correl} we show that most of the SMC stars seem to follow a similar relation as the Galactic OB stars. 
 
 Given the small sample of stars and the relatively large error bars in the derived stellar parameters  \citep[especially \logg, which can have error bars as high as 0.3 dex, ][]{2005Dufton}, we are limited in the interpretation of these results. However, a similar correlation (within the observational uncertainties of $\approx 20 \%$ of \vmac) between \vmac\ and \Ptmax in SMC further supports the connection between turbulent convection and macroturbulence. 

These SMC stars seem to follow a similar relation as the stars in the Galaxy, which suggests that a common relation exists between \vmac\ and \Ptmax/P from the FeCZ that is independent of the metallicity. To avoid confusion, we remark that the turbulent pressure does depend on the metallicity, as previously shown, while the relation between \vmac\ and \Ptmax might not. We also
note that the zero point of the linear relation at $\approx 30\,$km/s is due to the effects of microturbulence, given that this broadening mechanism (typically of about 10 km/s) is not taken into account separately when the macroturbulent velocities are derived. This is illustrated by \citet{2014SimonDiaz}, as microturbulence leads to an overestimate of the macroturbulent velocities especially when the relative contribution from microturbulence is high. Microturbulent velocities of about 10 km/s can lead to the erroneous interpretation of a macroturbulent velocity of about 15-40 km/s even in cases in which no macroturbulent broadening is present \citep{2014SimonDiaz}.

\begin{figure}
\resizebox{1.05\hsize}{!}{\includegraphics{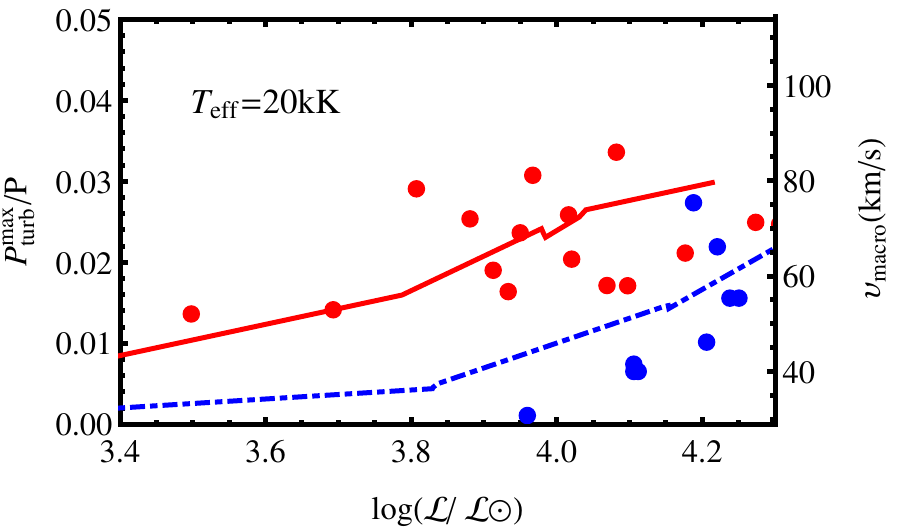}}
\caption{\vmac\ as a function of log($\mathcal{L}/\mathcal{L}_\odot$), with $\mathcal{L} := \Teff^4/g$, for the sub-set of SMC B-type stars from \citet{2006Dufton} (blue circles) and Galactic B-type stars from \citet{2015SimonDiaz} (red circles) in the effective temperature range \Teff $\approx 17-23 \,$kK, superposed on the computed \Ptmax/P (red continuous line for the MW, blue dashed for the SMC) as a function of log($\mathcal{L}/\mathcal{L}_\odot$) at \Teff =20kK.  }\label{corL}
\end{figure}

We furthermore compared the macroturbulent velocities of a sub-sample of the B SMC stars in \citet{2006Dufton} to that of some of the Galactic B stars in \citet{2015SimonDiaz} as a function of $\mathcal{L} := \Teff^4/g$ \citep{2014Langer} and constant effective temperature. We chose the temperature range where most of the SMC stars are located, which is in the range $\Teff\rm=17-23\,kK$ \citep[see also Fig. 2 by ][]{2006Dufton}, and thus restrict the samples in \citet{2006Dufton} and \citet{2015SimonDiaz} to the stars in this effective temperature range.  Based on Fig.\ref{ptL}, we predict that the stars in the Magellanic Clouds in general show lower values of \vmac\ than the stars in the Milky Way at fixed luminosity (and also for the same $\mathcal{L}$). This is indeed the case in Fig.\ref{corL}, where the SMC B stars on
average show lower \vmac\ values than the Galactic B stars in the same temperature range. 
Precise and systematic observations of a larger sample of stars in the Magellanic Clouds would make a good test case for these hypotheses.

\section{Conclusions}

We computed a set of stellar evolution models with SMC and LMC metallicity to compare the fraction of the pressure that is due to convective turbulence within the stellar envelope to the Galactic case, in different regions of the HR-diagram. We found a trend of lower turbulent pressure fractions for lower metallicities in the OB stars, with contributions of up to $\approx 3-4\%$ of the total pressure at $\logLl\approx 6$. 
This is due to the lower opacities in the sub-surface convective zones that are induced by the recombination of iron and iron-group elements in the SMC and LMC stellar models. Lower opacities lead to higher densities in the inflated envelopes, which also imply lower degrees of super-adiabaticity and thus lower convective velocities and lower turbulent pressure.   
No significant differences with the MW models are found in cool supergiants, for which the highest contribution from the turbulent pressure arises in the HCZ, accounting for up to 30\% of the total pressure.

The highly significant correlation between turbulent pressure and macroturbulent velocities for the Galactic stars found by \citet{2015GrassitelliA,2015GrassitelliB} has been interpreted considering that turbulent pressure fluctuations at the percent level 
may trigger high-order high-angular degree oscillations. It has
been shown that a large number of high-order pulsations can collectively mimic the effect of macroturbulence on the spectral lines by generating a velocity field at the surface on a scale larger than the size of the line-forming region \citep{2009Aerts}. 

Pulsations triggered by stochastic turbulent pressure fluctuations in the sub-surface convective zones, possibly together with classical $\kappa$-mechanism pulsations and strange mode pulsations, appear the only candidates at the moment able to explain the macroturbulent velocity fields observed in Galactic stars throughout the HR-diagram. Moreover, \citet{2015GrassitelliB} investigated a possible connection with some pulsating class of stars, namely the $\gamma$ Doradus class of pulsators, given that the location of the observational instability strip lies within the band (similar to those in Figs.\ref{LMCtr} and \ref{SMCtr}) of high contribution from turbulent pressure in the HCZ, that is, where $\Ptmax \geq 10\%$. Thus they suggested detailed asteroseismic calculations that include the effects of possibly also finite-amplitude pressure fluctuations associated with time-dependent turbulent convection \citep{2004Dupret,2014Antoci,2015GrassitelliB}.

Based on the results for the MW stars in the hot part of the HR-diagram, we predict that the Magellanic Clouds in general show lower macroturbulent velocities than the Galactic stars, as the turbulent pressure fraction is smaller for lower metallicities. This is supported by observational evidence in the results of \citet{2009Penny}, who found macroturbulent velocities to be
a function of metallicity in the O-star regime, given the lower \vsini\ found in the Magellanic Clouds. It is also supported by results for the SMC B-type stars by \citet{2006Dufton}, which are found to have lower macroturbulent velocities than the Galactic stars by \citet{2015SimonDiaz}. 

When we directly related the macroturbulent velocities from the SMC B-type stars by \citet{2006Dufton} to the corresponding fraction of turbulent pressure in the envelope of our stellar models, the two quantities appeared to be linearly related, as in \citet{2015GrassitelliA,2015GrassitelliB}, although with a lower statistical relevance of the result than
in the previous two cases. However, based on the results from the Galactic stars, the predicted macroturbulent velocities match observations fairly well (see Fig.\ref{corL}) and seem to follow the same relation as was found in the Galactic stars (see Fig.\ref{correl}). The trend of decreasing macroturbulent velocities in the effective temperatures range 10--30 kK found in the Galaxy and SMC by \citet{2010SimonDiaz}, \citet{2010Fraser}, and \citet{2006Dufton} also agrees well with our predictions. To interpret this trend, we note that as the stars evolve, the iron convective zone moves deeper inside the star, thus yielding to more efficient convection and therefore lower convective velocities.

We conclude that evidence from this and previous works lead to connect the origin of the macroturbulent broadening to the inefficiency of convection and the consequent vigorous turbulent motion in the sub-surface convective zones. Although the turbulent pressure arising from this motion does not appear to have a significant effect on the structure of 1D stellar models, it is intuitive to imagine that local stochastic pressure fluctuations in the partial ionization zones at the base of the envelope might induce perturbations of the hydrostatic equilibrium, which leads to variability at the surface \citep[e.g.,][]{2001Stein,2011Bedding}. Radiation-pressure-dominated inflated envelopes close to the Eddington limit and inefficient convection are strictly related \citep{2012Langer,2015Sanyal,2015Owocki}, thus macroturbulence together with asteroseismic calculations might be used to infer the conditions deep inside the envelopes of massive stars such as the strength and extent of inflation \citep{2015Sanyal} or the mixing-length parameter $\alpha$ \citep{2014Castro,2015GrassitelliB}. Additionally, inefficient convection might influence the wind of massive stars at its base, as the perturbations that originate in the convective zones might induce the formation of structures at or above the photosphere, especially for transonic surface velocity fields \citep{1984deJager,1996Cranmer,2002Prinja,2016Grassitelli}. Given the interplay between large-scale magnetic fields and convection, it is possible that strongly magnetic stars do not show vigorous convection, as the magnetic fields may dampen or inhibit convection. Therefore we expect that magnetic stars might be characterized by low macroturbulent velocities, as already noted by \citet{2013Sundqvist}.

\begin{acknowledgements} L.G. is part of the International Max Planck Research School (IMPRS), Max-Planck-Institut f{\"u}r Radioastronomie and University of Bonn and Cologne. L.F. acknowledges financial support from the Alexander von Humboldt foundation. L.G. thanks J.-C.Passy and O.Ram\'irez-Agudelo for valuable discussions and comments on this manuscript. 
\end{acknowledgements}


\end{document}